\documentclass[fleqn,a4paper]{vch-book}
\usepackage{amsmath}
\usepackage{amssymb}
\usepackage{makeidx}
    \makeindex
\usepackage{times}
\usepackage{tabularx}
\usepackage{booktabs}
\usepackage[dvips]{epsfig}
\DeclareGraphicsExtensions{.eps,.ps}

\chardef\bslash=`\\ 

\newcommand{\bibtex}{\ifx\is@itshape\f@shape{\fontshape{scit}\selectfont
Bib}\else\textsc{Bib}\fi\kern-.1em\TeX}

\begin{document}

\author{S.N. Dorogovtsev and J.F.F. Mendes}
\title{Accelerated growth of networks}


\maketitle

\begin{preface}\label{preface}
In many real growing networks the mean number of connections per vertex increases with time. The Internet, the Word Wide Web, collaborations networks, and many others display  this behavior. 
Such a growth can be called {\em accelerated}. We show that this acceleration influences distribution of connections and may determine the structure of a network. We discuss general consequences of the acceleration and demonstrate its features applying simple illustrating examples. In particular, we show that the accelerated growth fairly well explains the structure of the Word Web (the network of interacting words of human language). Also, we use the models of the accelerated growth of networks to describe a wealth condensation transition in evolving societies.      
\end{preface}



\section{Acceleration}\label{s-introduction} 

The great majority of models of evolving networks contain a very important assumption. These models suppose that the total number of edges in a growing network is a linear function of its size, that is, of the total number of vertices. This linear growth does not change the average degree of the network \cite{s01a,ab01a,dm01e}. (Here, following standard terminology from graph theory, we call the total number of connections of a vertex its degree. Physicists often call this quantity ``connectivity''. 
The number of incoming edges of a vertex in directed networks is called in-degree, the number of outgoing edges is out-degree.) 

The first model for the growth of networks under mechanism of preferential linking, namely, the Barab\'asi-Albert model \cite{ba99} (see also Ref. \cite{baj99}), is only one example of a linearly growing network from a very long list \cite{krl00,dms00,ab00a,dm00a,dm00b,kr00c,krr00a,dms01b}. Thus, a linear type of growth is usually supposed to be a natural feature of growing networks. But let us ask ourselves, whether this very particular case, that is, the linear growth is so widespread in real networks. 
To answer this question we must look at existing empirical data. 
Let us start from the most well known nets\vspace{7pt}. 

(i) {\em The World Wide Web:} 

Recall that the WWW is 
the array of its documents (pages) plus hyper-links, namely, mutual references in these documents. The WWW is a directed network. 
Although hyper-links are directed, pairs of counter-links, in principle, may produce undirected connections. 
Links inside pages (self-references) are usually not considered as edges of the WWW, so this network does not contain ``tadpoles'' (closed one-edge loops).  

According to Ref. \cite{bkm00}, in May of 1999, from the point of view of Altavista, the WWW consisted of $203\times 10^6$ vertices (URLs, i.e., pages) and $1466\times10^6$ hyperlinks. 
The average in- and out- degree were $\overline{k}_i=\overline{k}_o=7.22$. 
The average in- and out-degrees are equal to each other, since all the connections are inside the WWW. 
(Notice that ``physical'' time is unimportant for us, so that, in principle, we might not mention any date.) 
In October of 1999 there were already $271\times 10^6$ and $2130\times10^6$ hyperlinks. The average in- and out- degree have become $\overline{k}_i=\overline{k}_o=7.85$. Thus, the average degree of the WWW is increasing\vspace{7pt}. 

(ii) {\em The Internet:} 

Very roughly speaking, the Internet is a set of vertices, which are interconnected by wires. The vertices of the Internet are 
hosts (computers of users), servers (computers or programs providing 
a network service that also may be hosts), and routers that arrange traffic across the Internet. Connections are naturally undirected (an undirected network), 
and traffic (including its direction) changes all the time. 
Web documents are accessible 
through the Internet (wires and hardware), and this determines the relation between the Internet and the WWW. 
Routers are united in domains, however, this notion is not well defined for the Internet. In January of 2001, the Internet contained already 
about $100$ millions hosts. One should emphasize, that it is not the hosts that determine the structure of the Internet, but rather, routers and domains. In July of 2000, there were about $150\,000$ routers in the Internet \cite{gt00c}. Later, the number rose to $228\,265$ 
(data from Ref. \cite{yjb01a}).
Thus, one can consider the topology of the Internet on a router level or inter-domain topology \cite{fff99}. In the latter case, it is actually a small network. 

According to data of Ref. \cite{fff99} for the inter-domain level of the Internet, in November of 1997 it consisted of $3015$ vertices and $5156$ edges, so that the average degree was $\overline{k}=3.42$.  
In April of 1998 there were $3530$ vertices and $6432$ edges, and the average degree was $\overline{k}=3.65$. 
In December of 1998 there were $4389$ and $8256$ edges, so the average degree was already equal to $3.76$. Then, the average degree of the Internet on the inter-domain level is increasing. 

We have noted that domains in the Internet are poorly defined. Also, the last data of Ref. \cite{fff99} are for December of 1998. However, one may use more recent data on ``autonomous systems''. 
Extensive data on connections of operating ``autonomous systems'' (AS) in the Internet are being collected by the National Laboratory for Applied Network Research (NLANR). For nearly each day, starting from November of 1997, NLANR has a map of connections of AS. These maps are closely related to the Internet graph on the inter-domain level. Statistical analysis of these data was made in Ref. \cite{pvv01a,vpv01b}. The data were averaged, and for 1997 the average degree 
$3.47$ was obtained; in 1998, the average degree was $3.62$, in 1999, $\overline{k}=3.82$. Again we see that the average degree of the Internet on the inter-domain level (more rigorously speaking, on the AS level) is increasing.  One should add that the growth of the average degree of the net of AS was also indicated in Ref. \cite{gkk01e}.

Unfortunately, there are no reliable empirical data on the router level of the Internet to arrive at precise conclusions. 
In 1995, the Internet included 3888 routers with 5012 interconnections \cite{fff99}, that is $\overline{k} \sim 2.6$. 
In 2000, there were $\sim 150\,000$ routers and $\sim 200\,000$ interconnections between them, so that $\overline{k} \sim 2.7$ \cite{gt00c}. These data are taken from different sources, they are not precise 
 and cannot be compared\vspace{7pt}. 
  
(iii) {\em Networks of citations in scientific literature:} 

Vertices of citation networks are scientific papers, directed edges are citations. One cannot update the list of references in a published paper, so that new edges do not emerge between old papers. The direction of an edge between two papers is rigorously determined by their ages, so that one may forget about the directedness of citation networks. Such citation graphs (see Fig. \ref{f0}) are actually very simple growing networks, and most of demonstrating models of growing networks belong to this class. Note that in electronic archives one can update old papers and lists of references in them. This produce new links between old papers, so that the networks of citations of electronic archives are not quite classical citation graphs.      


\begin{vchfigure}[htb]
\includegraphics[scale=0.8]{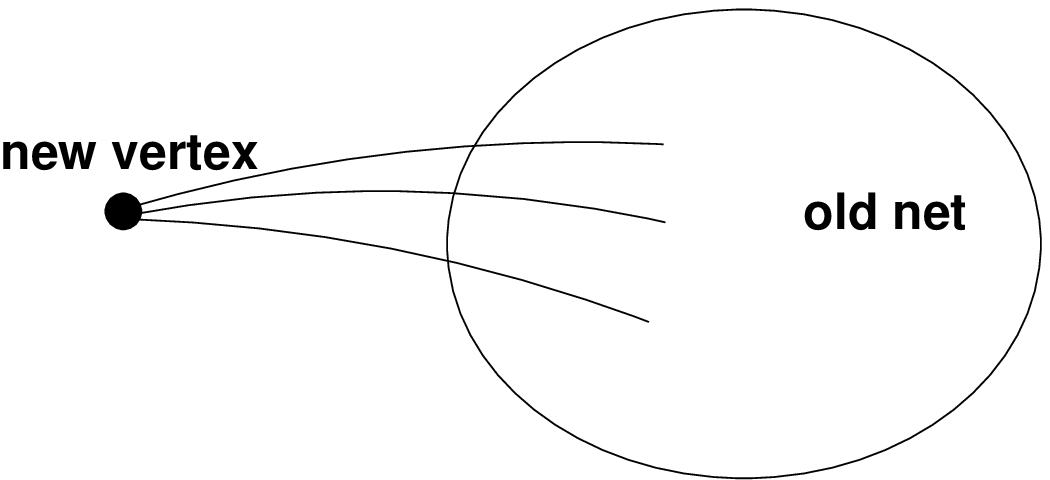}
         \vchcaption{Scheme of the growth of a citation graph. New connections emerge only between a new 
vertex and old ones. New connections between old vertices are impossible. 
}
         \label{f0}
\end{vchfigure}


Statistics of citations in scientific journals was studied in Ref. \cite{v01a} (see the earlier empirical study of the issue in Ref. \cite{r98}). These data were collected for a number of journals (about 10) in the period 1991-1999. 
In all the journals that were studied in Ref. \cite{v01a} the average number of references in papers was found to increase\vspace{7pt}.

(iv) {\em Collaboration networks:} 

In the simplest version of a collaboration network, vertices are collaborators. A pair of vertices is connected together by an undirected edge if there was at least one act of collaboration between them \cite{ba99,asbs00}. For example, in scientific collaboration networks (networks of coauthorships), vertices are authors, 
and edges are coauthorships \cite{n00}. Such networks are projections of more complex and informative bipartite graphs, which contains of two types of vertices: collaborators and acts of collaboration. Each collaborator is connected to all the acts of collaboration, in which he was involved. Empirical data are mostly collected for simple one-mode collaboration networks. 
 
Empirical data of Refs. \cite{bjnr01a,jnb01a} for large scientific collaboration networks indicate the linear growth of their average degree with the increasing number of their vertices. This means that the total number of edges in a network increases as a square of the total number of vertices. 

Thus we see that the accelerated growth of networks is not an exception but rather a rule. On the contrary, the linear growth is a simple but very particular case.


\section{Reasons for the acceleration}\label{s-degree}

Why is the accelerated growth widespread? As an example consider the growth of the WWW. Let us discuss how new pages appear in the WWW (see Fig. \ref{f00}) \cite{dm01e}. 
Discussion of the growth of the WWW may be found in Refs. \cite{be00,dms00f}.
Suppose, you want to create your own personal home page. You prepare it, put references to some pages of the WWW (usually, there are several such references, but in principle the references may be absent), etc. But this is only the first step. You must make your page accessible in the WWW. Your system administrator puts a reference to it (usually one reference) in the home page of your institution, and your page in the Web. 


\begin{vchfigure}[htb]
\includegraphics[scale=0.8]{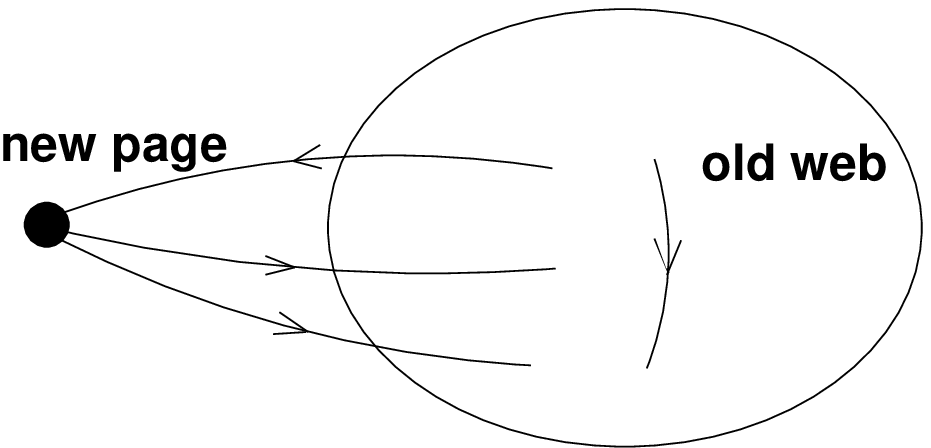}
         \vchcaption{Scheme of the growth of the WWW. A new document of the WWW must have at least one incoming hyperlink to become accessible. It may contain any number of references to other pages of the WWW, but usually there are several such outgoing hyperlinks. Also, new hyperlinks emerge between old pages of the WWW.
}
         \label{f00}
\end{vchfigure}


However, you proceed to work with your page. From time to time, you add new references to it. Of course, you may remove some old references, but usually the total number of references in a page grows. Then the average degree of the WWW increases, that is, the growth of the WWW is naturally accelerated.


\section{Degree distributions of networks}\label{s-degree}


\subsection{Types of degree distribution}\label{ss-types}

In this paper we restrict ourselves to degree distributions of networks. Most of empirical results are obtained for this simple basic characteristic. 
Unfortunately, a degree distribution (in-, out-degree distribution) is a restricted characteristic of networks. Indeed, degree is a one-vertex quantity, so that, in general, degree distribution does not yield information about the global topology of a network. 

In most of cases, for example, for growing networks, in which correlations between degrees of vertices are strong \cite{kr00c,pvv01a,vpv01b}, a degree distribution is only the tip of the iceberg (see Fig. \ref{f1}, a). Of course, if degree-degree correlations in a network are absent, then, knowing the degree distribution of a network, one can completely characterize the net (see Fig. \ref{f1}, b). 
We face this situation in many equilibrium networks.  


\begin{vchfigure}[htb]
\includegraphics[scale=0.6]{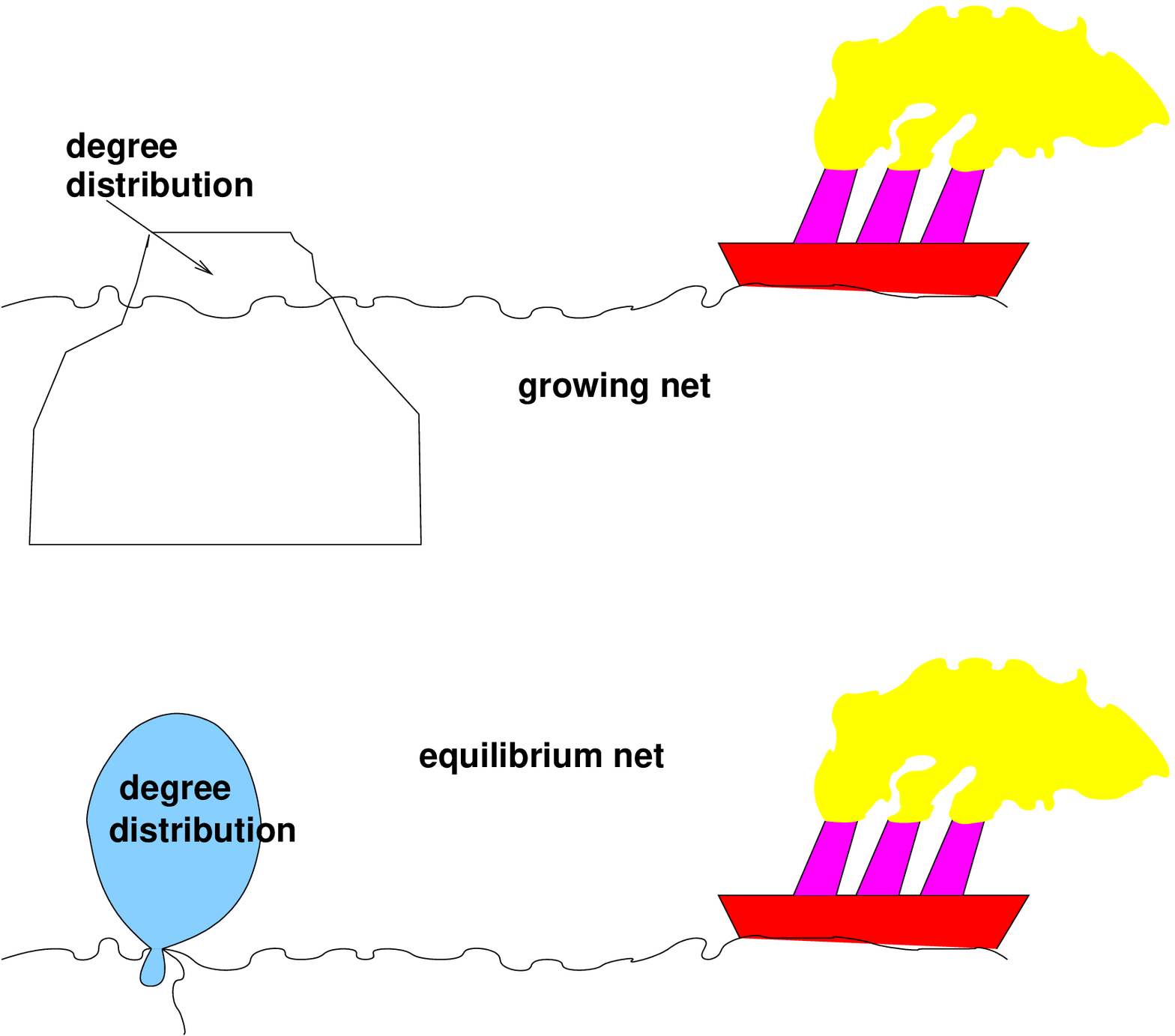}
         \vchcaption{Degree-degree correlations, which are necessary 
present in growing networks, make a degree distribution far less informative characteristic (a). 
The degree distribution of the equilibrium uncorrelated network contains 
complete information about its structure (b). 
}
         \label{f1}
\end{vchfigure}


Furthermore, analytical results on percolation on networks \cite{mr95,nsw00}, disease spread within them \cite{pv00,pv01}, etc. were obtained just for a simple construction without degree-degree correlations. This construction is a standard model of a maximally random graph with an arbitrary degree distribution taken from mathematical graph theory (``random graphs with restricted degree sequences'') \cite{b80}. Luckily, it seems that main percolation and disease spread results that was obtained for equilibrium networks are still valid for non-equilibrium nets. 

What kinds of  degree distributions are realized in networks? 
Here we list the main types with some simple examples of the corresponding networks. 

(a) Poisson degree distribution, 
$P(k) = e^{-\overline{k}}\overline{k}^k/k!$ (see Fig. \ref{f2}, a). 

The Poisson distribution is realized in a classical random equilibrium graph of Erd\"os and R\'enyi \cite{er59,er60} in the limit of the infinite network, that is, when the total number of vertices $N$ is infinite. 
Pairs of randomly chosen vertices are connected by edges. Multiple edges (``melons'') are forbidden. One may create $L$ edges in the graph, or connect pairs of vertices with the probability $L/[N(N-1)/2]$. In both these cases, the resulting graph is the same in the limit $N \to \infty$. 


\begin{vchfigure}[htb]
\includegraphics[scale=0.6]{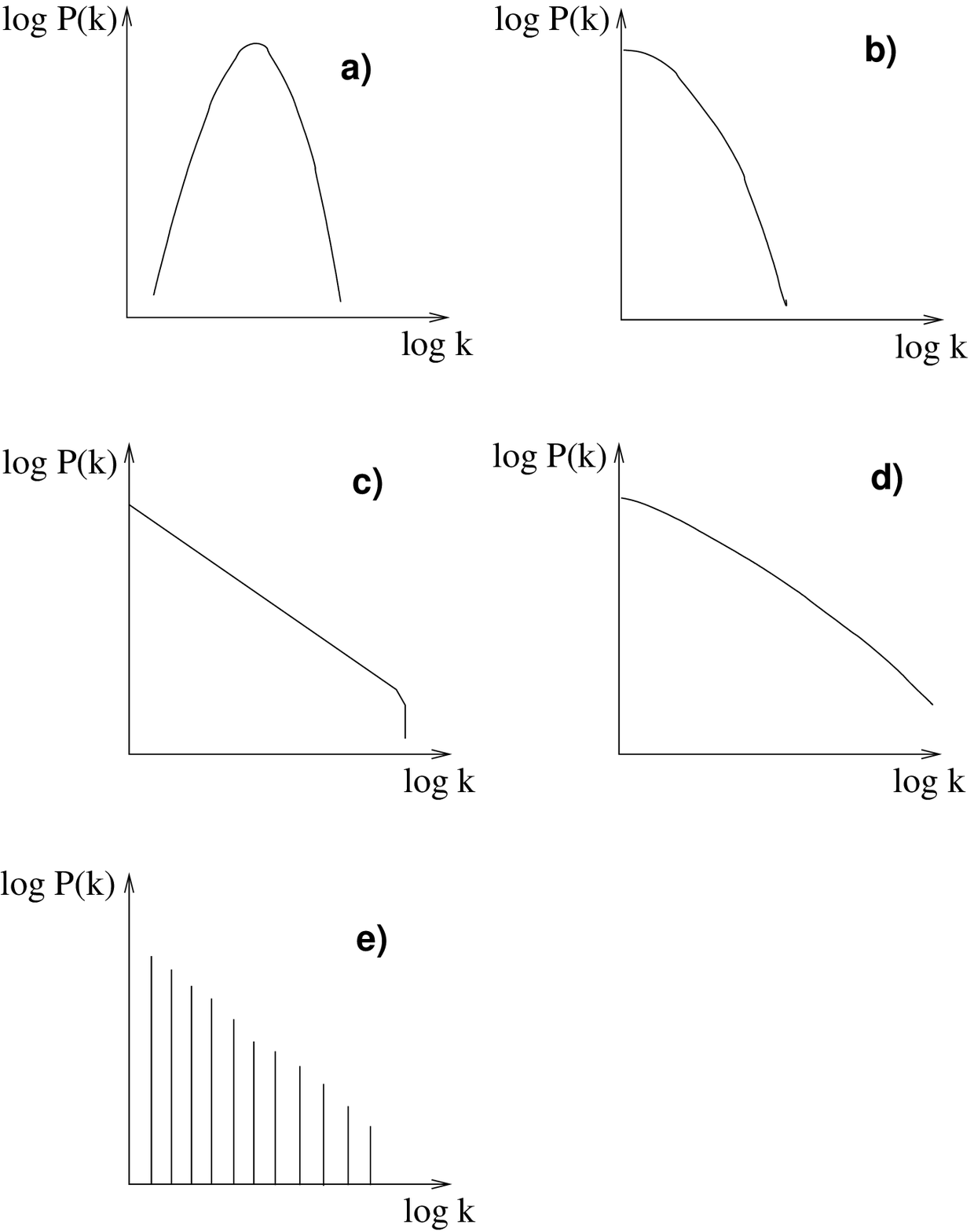}
         \vchcaption{``Zoology'' of degree distributions in networks. 
Main types of a degree distribution in log-log plots. 
Poisson (a), exponential (b), power-law (c), multifractal (d), 
and discrete (e) distributions.  
}
         \label{f2}
\end{vchfigure}


(b) Exponential degree distribution, 
$P(k) \sim \exp(-k/\mbox{const})$ (see Fig. \ref{f2}, b). 

A citation graph (see Fig. \ref{f0}) with attachment of new vertices to randomly chosen old ones produces the exponential distribution, but this is only one possible example. (Let each new vertex have the same number of connections, that is, the growth is linear.)

Also, the exponential degree distribution is rather usual for many equilibrium networks that are constructed by mechanism of preferential linking.

(c) Power-law degree distribution, 
$P(k) \sim k^{-\gamma}$ (see Fig. \ref{f2}, c). 

Here the standard example is the Barab\'asi-Albert model \cite{ba99} (see also Ref. \cite{baj99}). This growing network is a linearly growing citation graph in that new vertices are attached to preferentially chosen old ones. ``Popular'' old vertices attract more new connections than ``failures'': ``{\em popularity is attractive}''. 
This is a quite general principle. For example, this one is incorporated in the Simon model \cite{s55,s57}. 
In the Barab\'asi-Albert model, the probability that an edge becomes attached to some vertex is proportional to the degree $k$ of this vertex. This yields $\gamma=3$. If the probability is proportional to $k+\mbox{const}$ (a linear preference function), $\gamma$ takes values between $2$ and $\infty$ 
as the constant changes from $-1$ to $\infty$ \cite{dms00}. 

Power-law distributions are usually called scale-free or fractal.

(d) Multifractal degree distributions (see Fig. \ref{f2}, c). 

This distribution has a continuum spectrum of power laws with different weights. The growth of a network may produce such a degree distribution if new vertices partially copy degrees of old ones \cite{dms02}. In particular, multifractal degree distributions emerge in some models of networks of protein-protein interactions \cite{vfmv01a}. Multifractal distributions is a more general case of a fat-tailed distribution than power-law distributions. 
Numerous empirical data were fitted by a power-law dependence. However, 
there were no attempts to check the possibility that at least some of empirical degree distribution are multifractal. 

(e) Discrete degree distributions (see Fig. \ref{f2}, d). 

Deterministic growing graphs have a discrete spectrum of degrees. Recently, it was demonstrated that some simple rules of deterministic growth may produce discrete degree distributions with a power-law decay \cite{brv01a}. 
Moreover, deterministic graphs from Refs. \cite{dm01e,dgm01,jkk01d} have an average shortest-path length, which is proportional to the logarithm of their size. Figure \ref{f3} shows a simple deterministic graph \cite{dm01e,dgm01} with the discrete degree distribution that is characterized by exponent $\gamma = 1 + \ln3/\ln2$.    


\begin{vchfigure}[htb]
\includegraphics[scale=0.5]{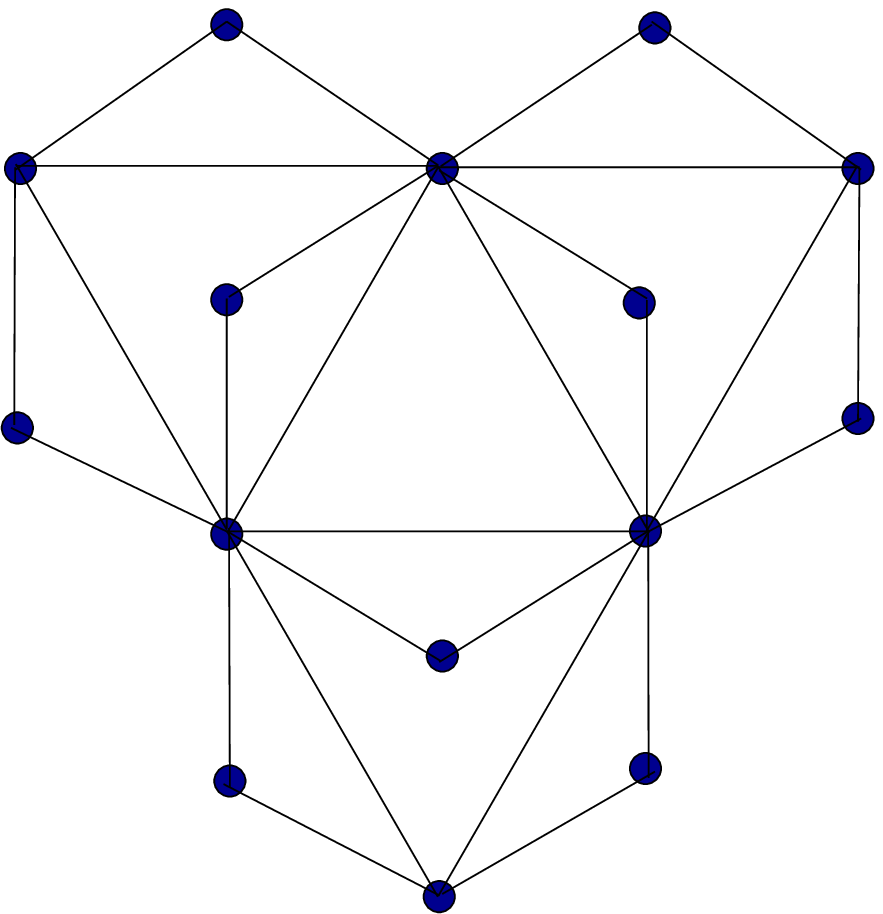}
         \vchcaption{A simple deterministic graph \protect\cite{dm01e,dgm01} with a power-law 
discrete degree distribution. 
The growth starts from a single edge between two vertices. At each time step, each edge of the 
graph generates a new vertex, which became attached to both the end vertices of the mother edge.
The average shortest-path length 
of this graph grows logarithmically with the total number of vertices.
}
         \label{f3}
\end{vchfigure}



\subsection{Power-law degree distribution}\label{ss-power}

Power-law (that is, ``scale-free'') degree distributions is a prominent particular case of fat-tailed degree distributions, which are widespread in real networks (both natural and artificial) \cite{ba99,baj99,asbs00}.  
Let us discuss briefly the general features of power-law distributions. 

One may ask, what are the possible values for $\gamma$? 
The first natural restriction follows from from the normalization condition $\int dk P(k) = 1$ (in this discussion we change the corresponding sum to the integral). We may not be worried about the low-degree region, since the degree distribution is certainly restricted below some characteristic degree $k_0$. Only the large degree behavior of the degree distribution is interesting for us.  Therefore, the strong restriction is $\gamma>1$, otherwise the integral is divergent. 

If a network grows linearly, so that the first moment of the distribution, that is, the average degree $\overline{k}$, is independent of time, then we have the second restriction $\int dk k P(k) < \infty$. Therefore, $\gamma>2$ for linearly growing networks. 

Finite size effect cuts the power-law part of the degree distribution at large degrees. This produces size-dependent degree distributions.  
One may easily estimate the position of the cutoff $k_{cut}$ in the situation where $\gamma>2$. 
Let the total number of vertices in the net be $t$, and $k_0$ be some characteristic degree, below which the distribution is, for example, constant or even zero. 
Then, using the normalization $\int dk P(k) = 1$ gives the power-law part of the degree distribution of the form 
$P(k) \sim [(\gamma-1)k_0^{\gamma-1}] k^{-\gamma}$ for $k_0 < k < k_{cut}$. 

When one measures the degree distribution of a network using only one realization 
of the growth process, strong fluctuations are observed at degree $k_f(t)$ that is determined by the condition $t P(k_f(t)) \sim 1$. This means that only one vertex in the network has such degree. (More rigorously speaking, the number of such vertices is of the order of one.)
This is the first natural scale of the degree distribution. 

One may improve the statistics by measuring many realizations of the growth process, or, for example, by passing to the cumulative distribution $P_{cum} \equiv \int_k^\infty dk\,P(k)$. Both these tricks allow us to reduce the above fluctuations. However, we still cannot surpass the next threshold that is originated from the second natural scale, $k_{cut}$: $t P_{cum} (k_{cut}(t)) \sim 1$. This means that only one vertex in the network is of degree greater than $k_{cut}$. 
(Again, more rigorously, the number of such vertices is of 
the order of one.) Using the above expression for $P(k)$ gives 

\begin{equation}
\label{1}
k_{cut} \sim  k_0 \, t^{1/(\gamma-1)} 
\, .  
\end{equation} 

Notice that the only reason for this estimate for the cutoff is the natural scale of the problem. Hence more convincing arguments are necessary. The estimate was checked for some specific models. A growing network \cite{dms01b} was solved exactly, and the exact position of the cutoff have coincided with Eq. (\ref{1}). The degree distribution of this network has a typical form (see Fig. \ref{f4}). Notice a hump near $k_{cut}$ in Fig. \ref{f4}. This is a trace of initial conditions. 
Simulation of a scale-free equilibrium network \cite{bck01} also yielded the cutoff at this point. However, the introduction of the death of vertices in the network may change the estimate (\ref{1}). 
This factor also removes the hump from the degree distribution. 
Here we do not consider such situations. 


\begin{vchfigure}[htb]
\includegraphics[scale=0.8]{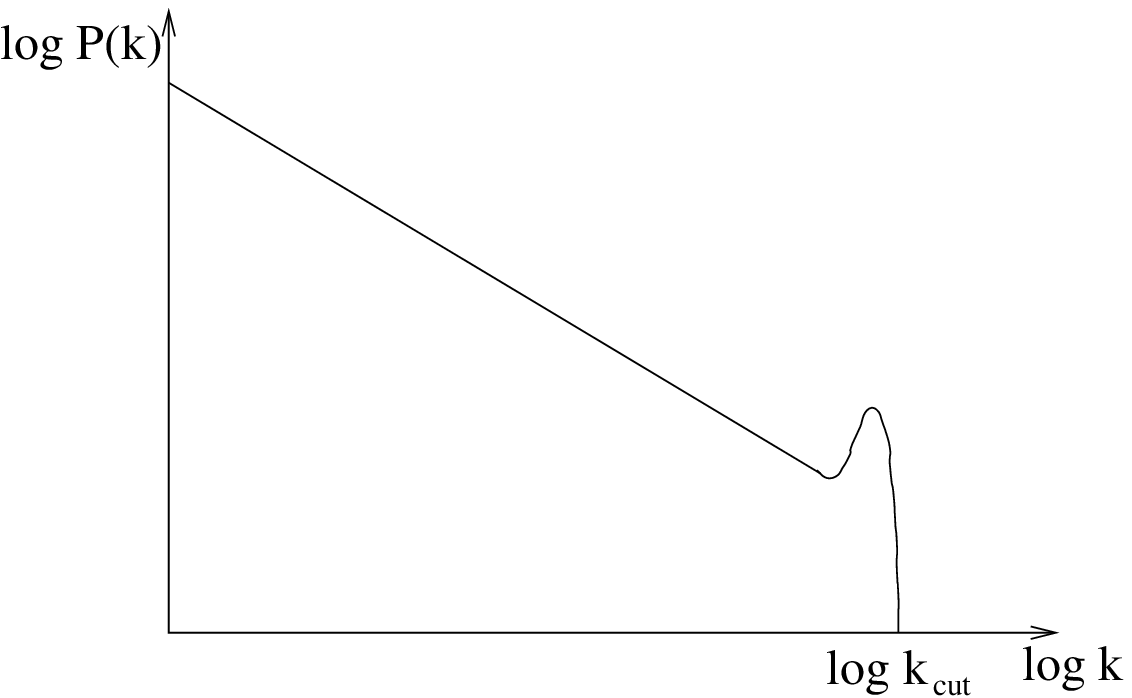}
         \vchcaption{The typical form of a power-law degree distribution of finite 
growing networks. The finite-size cutoff is given by Eq. (\protect\ref{1}). 
The hump near the cutoff depends on initial conditions (we do not account for the factor of mortality).
}
         \label{f4}
\end{vchfigure}


The cutoff (\ref{1}) hinders measurements of power-law dependences in networks \cite{dms01b}. From Eq. (\ref{1}) one sees that the measurements of large enough $\gamma$ are actually impossible. Indeed, in this case $k_{cut}$ is small even for very large networks, and there is no room $\ln k_0 < \ln k < \ln k_{cut}$ for fitting.   

What is the nature of power-laws in networks? 
One may directly relate them to self-organized criticality. 
While growing under mechanism of preferential linking, networks self-organize into scale-free structures, that is, are in a critical state. This critical state is realized for a wide range of parameters of preferential linking, namely for any linear preference function (more rigorously, for any preference function which is asymptotically linear at large $k$ \cite{krl00}). The linear growth of networks may produce scale-free structures. Then, one may ask: 
What degree distributions does the accelerated growth produce?


\section{General relations for the accelerated growth}\label{s-general}

Let us start with general considerations and do not restrict ourselves by some specific model.  
Let the average degree grows as a power of $t$, $\overline{k} \propto t^a$, that is, the total number of edges $L(t) \propto t^{a+1}$. Here $a>0$ is the growth exponent. 
The consideration is valid not only for degree, but also for in-, and out-degrees, so we use the same notation $k$ for all them.  
The power-law type of acceleration we have chosen since one may hope that it provide scale-free networks. We suppose from the very beginning that this is the case and then check our assumption. 

For the accelerated growth, the degree distribution may be non-stationary. 
It is natural to choose its power-law part in the form 

\begin{equation}
\label{1a}
P(k,t) \sim t^z k^{-\gamma} 
\, .  
\end{equation}
Here we have introduced new exponent $z>0$ \cite{dm01e,dm01c,dm01d} (recall that we consider only $a>0$). This form is valid only in the range 
$k_0(t) < k < k_{cut}(t)$. 
Using the normalization condition $\int_{k_0(t)}^\infty dk\,t^z  k^{-\gamma} \sim 1$ 
gives 

\begin{equation}
\label{2}
k_0(t) \sim t^{z/(\gamma-1)} 
\, .  
\end{equation} 
This estimate is valid for any $\gamma>1$. 

The cutoff $k_{cut}(t)$ is estimated from the condition 
$t \int_{k_{cut}(t)}^\infty dk\, t^z k^{-\gamma} \sim 1$. Therefore, 

\begin{equation}
\label{3}
k_{cut}(t) \sim t^{(z+1)/(\gamma-1)}
\,   
\end{equation} 
(compare with Eq. (\ref{1}) for the linear growth.) Equation (\ref{3}) holds for any $\gamma>1$. 

We will consider two cases (see Fig. \ref{f5}), $1<\gamma<2$ and $\gamma>2$. Recall that we do not account for mortality of vertices. 


\begin{vchfigure}[htb]
\includegraphics[scale=0.8]{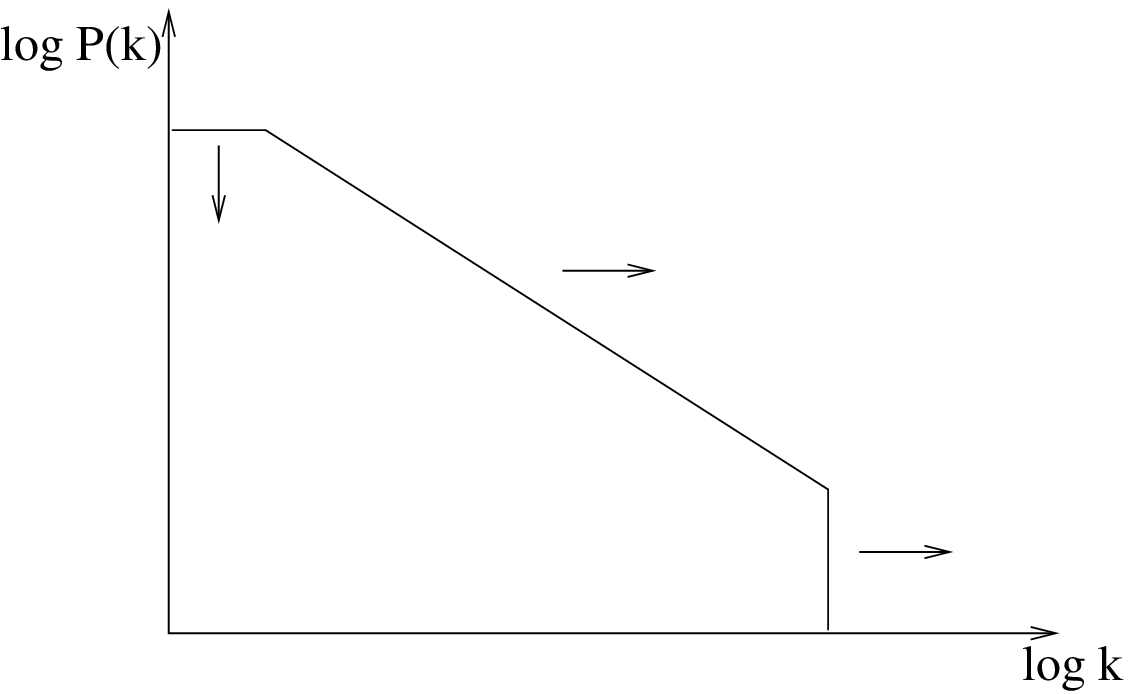}
         \vchcaption{Schematic plot of a time-dependent degree distribution of networks that grow in the accelerated mode. Arrows show how the degree distribution changes with time.
}
         \label{f5}
\end{vchfigure}


(i)  $1<\gamma<2$. 

Recall that the average degree distribution $\overline{k}(t) \sim t^a$. Then 
$$
t^a \sim \int^{t^{(z+1)/(\gamma-1)}}dk\, k t^z k^{-\gamma} \sim t^{-1+(z+1)/(\gamma-1)} 
\, .  
$$
Here the value of the integral is determined by its upper limit. 
Therefore, $(z+1)/(\gamma-1)=a+1$, and the cutoff is of the order of the total number of edges in the network, 

\begin{equation}
\label{3a}
k_{cut}(t) \sim t^{a+1} \sim L(t) 
\, .  
\end{equation} 
But this is the maximum possible degree in the problem. In this sense, any cutoff of a degree distribution is absent if $\gamma<2$. From the last relation, we obtain $\gamma$ exponent in such a situation, 

\begin{equation}
\label{4}
\gamma = 1 + \frac{z+1}{a+1}
\, .  
\end{equation} 
Here, for $\gamma<2$, one assumes that $z<a$. The lower boundary for $\gamma$, namely $\gamma=1+1/(a+1)$ is approached when $z=0$, that is, when the distribution is stationary\vspace{7pt}. 

(ii) $\gamma>2$. 

The integral for the average degree is determined by its lower limit 
$$ 
t^a \sim \int_{tz/(\gamma-1)} dk\, k t^z k^{-\gamma} \sim 
t^{z-z(\gamma-2)/(\gamma-1)} \, .
$$
Hence 

\begin{equation}
\label{5}
\gamma = 1 + \frac{z}{a}
\,   
\end{equation} 
and $z>a$ to keep $\gamma>2$. Notice that this relation is not valid for $a=0$. One sees that, in this case, the degree distribution cannot be stationary: $z>a>0$.


\section{Scaling relations for accelerated growth}\label{s-scaling}

For simple scale-free networks that grow in a linear mode, simple scaling relations can be written \cite{dms00,dm00a}. Let us briefly describe the corresponding scaling relations for the accelerated growth. 
If vertices in a growing network do not die, one can label them by their ``birth date'' $0<s<t$. We denote by $p(k,s,t)$ the probability that the vertex $s$ is of the degree $q$. The average degree of a vertex $s$ at time $t$ is 
$\overline{k}(s,t) \equiv \int dk\, k p(k,s,t)$. 

For networks that we consider the $\overline{k}(s,t)$ is 

\begin{equation}
\label{6}
\overline{k}(s,t) \propto t^\delta \left(\frac{s}{t}\right)^{-\beta}
\, ,  
\end{equation} 
where $\beta$ and $\gamma$ are scaling exponents. One can show \cite{dm01d} that $p(k,s,t) = [1/\overline{k}(s,t)] g[k/\overline{k}(s,t)]$, where $g[\ ]$ is some scaling function, therefore 

\begin{equation}
\label{7}
p(k,s,t) = t^{-\delta} \left(\frac{s}{t}\right)^\beta 
g \left[ k t^{-\delta} \left(\frac{s}{t}\right)^\beta \right]
\, .  
\end{equation}  
Using the relation $P(k,t) = t^{-1} \int_0^t ds p(k,s,t)$ yields 

\begin{equation}
\label{8}
\int_0^\infty dx\,t^{-\delta} x^\beta g[k t^{-\delta} x^\beta] \propto 
t^{\delta/\beta} k^{-1-1/\beta} \propto t^z k^{-\gamma}
\, ,  
\end{equation}  
whence we obtain relations for the scaling exponents: 

\begin{equation}
\label{9}
\gamma = 1+1/\beta
\,   
\end{equation}  
and

\begin{equation}
\label{10}
z = \delta/\beta
\, .  
\end{equation} 

Accounting these relations gives the scaling form

\begin{equation}
\label{11}
p(k,s,t) = \frac{s^{1/(\gamma-1)}}{t^{(z+1)/(\gamma-1)}} 
g \left[ k \frac{s^{1/(\gamma-1)}}{t^{(z+1)/(\gamma-1)}} \right]
\, .  
\end{equation} 
Similarly, one can find the scaling form for the degree distribution. 

\begin{equation}
\label{12}
P(k,t)  = t^z k^{-\gamma} G(k t^{-(1+z)\beta}) = 
t^z k^{-\gamma} G(k t^{-(1+z)/(\gamma-1)})
\, ,  
\end{equation} 
where $G(\ )$ is a scaling function. When $z=0$, Eqs. (\ref{11}) and (\ref{12}) coincide with the scaling relations \cite{dms00,dm00a} for linearly growing networks. 

Notice that it is sufficient to know $a$ and only one exponent of $\gamma,\beta,z,\delta$, or $x$ to find all the others.


\section{Degree distributions produced by the acceleration}\label{s-models}

Let us discuss several illustrative examples. To begin with, we consider a network growing under mechanism of preferential linking, in which number of new connections increases as a power law in time. 
At this point we do not discuss the origin of this power-law dependence.  
Let it be equal to $c_0 t^a$, where $c_0$ is some positive constant.  Here is convenient to study the in-degree distribution, so that $k$ will be in-degree. In such an event we are interested only in incoming connections, so that the outgoing ends of new edges may be attached to any vertices of the network or even be outside of the net. 

Let the probability that a new edge becomes attached to a vertex of in-degree $k$ be proportional to $k+A(t)$, where $A(t)$ is some additional attractiveness of vertices. Two particular cases of this linear preferential linking are considered below in the framework of a simple continuum approach \cite{baj99,dm00a,dm01d}.


\subsection{Model for $\gamma<2$}\label{ss-less}

If the additional attractiveness is constant, $A=\mbox{const}$, the continuum equation for the average in-degree $\overline{k}(s,t)$ of individual vertices that born at time $s$ and are observed at time $t$ is of the form 

\begin{equation}
\label{15}
\frac{\partial \overline{k}(s,t)}{\partial t} = 
c_0 t^a \frac{\overline{k}(s,t)+A}{\int_0^t du[\overline{k}(u,t)+A]} 
\,   
\end{equation}    
with additional starting and boundary conditions $\overline{k}(0,0)=0$ and $\overline{k}(t,t)=0$. Here we supposed that new vertices have no 
incoming edges. We use this assumption only for brevity. Naturally, the total in-degree of the network is 
$\int_0^t du \overline{k}(u,t) = c_0 t^{a+1}/(a+1)$. 
This also can be seen by integrating both the sides of Eq. (\ref{15}) over $s$. Accounting the last equality yields the solution of Eq. (\ref{15}): 

\begin{equation}
\label{16}
\overline{k}(s,t) = A \left(\frac{s}{t}\right)^{-(a+1)}
\, .  
\end{equation} 
Therefore, $\beta$ exponent equals $a+1>1$, so that using scaling relation (\ref{9}) gives 

\begin{equation}
\label{17}
\gamma = 1 + \frac{1}{a+1} < 2
\, .  
\end{equation}  

One may also apply the following simple relation of the continuum approach: 

\begin{equation}
\label{18}
P(k,t) = \frac{1}{t}\!\int_0^t \!ds\, \delta(k-\overline{k}(s,t)) = 
-\frac{1}{t}\! 
\left.\left( \frac{\partial\overline{k}(s,t)}{\partial s} \right)^{-1}\right|_{s=\overline{k}(s,t)} 
.  
\end{equation}  
This equality follows from the fact that the solution of the master equation for the probability $p(k,s,t)$ in the continuum approximation is the $\delta$-function. From Eqs. (\ref{16}) and (\ref{17}) we obtain the 
in-degree distribution 

\begin{equation}
\label{19}
P(k,t) = \frac{A^{1/(a+1)}}{a+1} k^{-[1+1/(a+1)]}
\,  ,
\end{equation}  
which is stationary. We have shown in Sec. \ref{s-general} that when $\gamma=1+1/(a+1)$, the (in-) degree distribution must be stationary, and exponent $z$ is zero. This is the case for the network under consideration.


\subsection{Model for $\gamma>2$}\label{ss-greater}

Now we choose a different rule of attachment of new edges to vertices. 
Let the additional attractiveness be time dependent. Furthermore, let it be 
proportional to the average in-degree of the network, $c_0 t^a/(a+1)$, at the birth of an edge, 
$A(t) = B c_0 t^a/(a+1))$. Here $B>0$ is some constant. Analogously to the above we obtain the non-stationary in-degree distribution 

\begin{equation}
\label{20}
P(k,t) \sim t^{a(1+B)/(1-Ba)} k^{-[1+(1+B)/(1-Ba)]}
\,  
\end{equation} 
for $k \gg t^a$. Hence $\gamma$ exponent is 

\begin{equation}
\label{21}
\gamma = 1 + \frac{1+B}{1-Ba} > 2
\, .  
\end{equation} 
The scaling regime is realized when $Ba<1$.


\subsection{Dynamically induced accelerated growth}\label{ss-dinamically}

We have shown above that the power-law growth of the total number of edges in a network (or its average degree) produces fat-tailed distributions. Now we discuss reasons the for the power-law growth. 

Consider an undirected citation graph, in which each new vertex becomes attached to a randomly chosen old one plus to some of its nearest neighbors, to each one of them with probability $p$ (see fig. \ref{f6}). For the total number of edges $L(t)$ one can write 

\begin{equation}
\label{22}
L(t+1) - L(t) = 1 + p \overline{k}(t)
\, .  
\end{equation} 
Here we use the continuum approximation. $\overline{k}(t) = 2L(t)/t$, therefore 

\begin{equation}
\label{23}
\frac{1}{2} \frac{d}{dt}[t\overline{k}(t)] = 1 + p \overline{k}(t)
\, .  
\end{equation} 

For $p<1/2$, the solution of this equation approaches the stationary limit $\overline{k}=2/(1-2p)$ as $t \to \infty$. In this case the degree distribution is stationary, and $\gamma$ exponent is $\gamma=1+1/p>3$. 


\begin{vchfigure}[htb]
\includegraphics[scale=0.8]{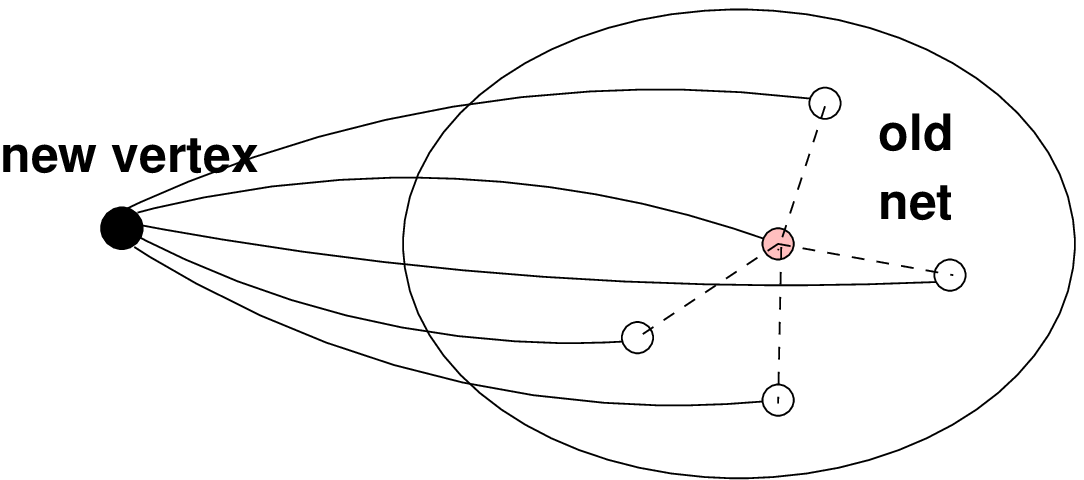}
         \vchcaption{One of possibilities to obtain the acceleration. 
In this citation graph, each new vertex becomes attached 
to a randomly chosen old one plus to some 
of its nearest neighbors.
}
         \label{f6}
\end{vchfigure}


The situation for $p>1/2$ is quite different, the average degree of the network growth as a power law, $\overline{k}(t) \sim t^{2p-1}$ for large networks. 
This produces non-stationary distribution $P(k) \propto t^{z}k^{-\gamma}$ with $\gamma=1+1/(1-p)>3$ and $z=1/(1-p)-2$.
Of course, other mechanisms for the accelerated growth are also possible.


\subsection{Partial copying of edges and multifractality}\label{ss-copying}

From Eq. (\ref{1}) for the cutoff of a power-law (or, which is the same, fractal) distribution, one sees that the size dependence of the moments 
$M_m(t) \equiv \int dk\, k^m P(k,t)$ of this distribution is   

\begin{equation}
\label{24}
M_m(t) \propto t^{\tau(m)}
\, ,  
\end{equation} 
where $\tau(m)$ exponent is a linear function of the order $m$ of a moment, 
$\tau(m) = (m-1)/(\gamma-1) - (\gamma-2)/(\gamma-1)$. Just the linearity of 
$\tau(m)$ defines a fractal distribution. 
The size dependence of the moments of a multifractal distribution also has the form (\ref{24}), but its 
$\tau(m)$ exponent is a nonlinear function of $m$. 

Multifractal distributions are a more general case of fat-tailed distributions than a power-law (fractal, scale-free) dependence. In Sec. \ref{ss-dinamically} we have shown how the accelerated growth may generate fractal distributions. 
However, this is only a particular possibility.   
Partial copying (partial inheritance) of degrees of old vertices by newborn ones 
together with the preferential attachment of some extra new edges  
usually provides networks, which grows in a nonlinear way and have multifractal degree distributions.  

A simple consideration of this problem can be found in Ref. \cite{dms02}. 
Note that the acceleration and the multifractality of the degree distribution were obtained in a similar model \cite{vfmv01a} for protein-protein interaction networks. In this model,  
duplication of vertices with edges attached to them and breaking of some connections of parent vertices were used  
instead of partial copying in Refs. \cite{dms02}.


\section{Evolution of the Word Web}\label{s-word}

The weak point of network science is the absence of a convincing comparison 
of numerous schematic models with real networks. Most of models of growing networks only demonstrate intriguing effects but, in fact, are very var from reality. Available empirical data usually can be explained by applying various models with fitting parameters. As a rule, only the exponent of the empirical degree distribution is used for comparison.   

Here we consider an exceptional situation, where a reasonable comparison of the model of a growing  network with empirical data is possible {\em without any fitting}.  
Moreover, it is the idea of the accelerated growth that yields an excellent agreement. 
 
The problem of human language is a matter of immense interest of various sciences. How did language begin? How does language evolve? What is its structure? 
Quite recently, a novel approach to language was proposed \cite{fs01}. 
Human language was considered as a complex network of interacting words. 
Vertices in this Word Web are distinct words of language, and undirected edges are connections between interacting words. 

Words interact when they meet in sentences. 
Different reasonable definitions yield very similar structures of the Word Web. For example, we can connect the nearest neighbors in sentences. This means that the edge between two words of language exists if these words are the nearest neighbors in at least one sentence in the bank of language. One sees that multiple connections are absent. 
Of course, this is a rather naive definition, but it is also possible to account for other types of correlations between words in a sentence \cite{fs01}. 
The resulting network gives the image of language, which is available for statistical analysis. 

The empirical degree distribution  \cite{fs01} of the Word Web is very complex (see Fig. \ref{f7}). Therefore, a perfect description of these data without fitting would be convincing. Indeed, it is hardly possible to describe such a complex form of the distribution completely by coincidence. We show below that a minimal model of the evolving Word Web \cite{dm01f}, with only known parameters of this network, provides such a perfect description.      


\begin{vchfigure}[htb]
\includegraphics[scale=0.4]{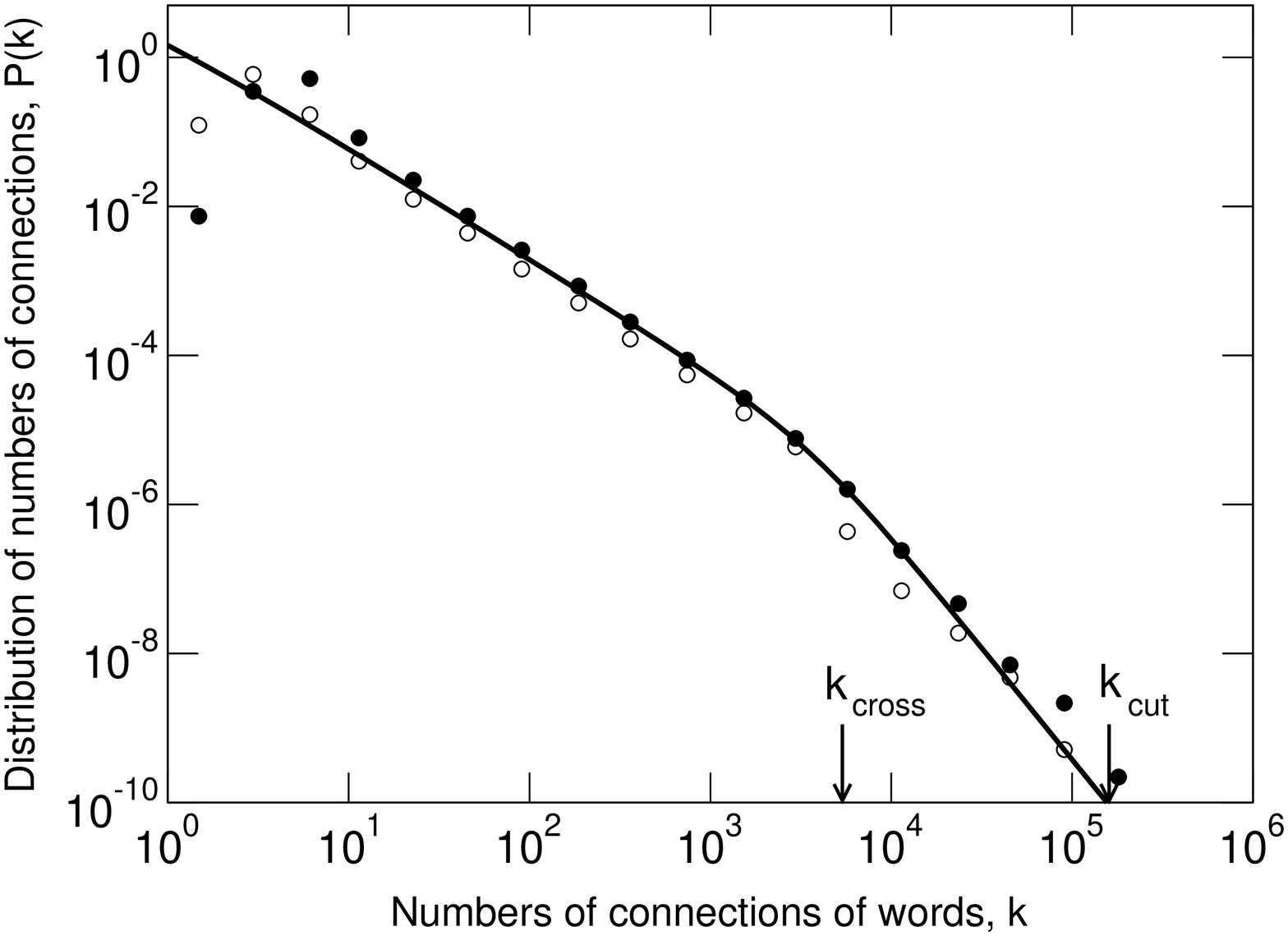}
	 \vchcaption{Empirical degree distribution of the Word Web (points) \protect\cite{fs01}. 
Empty and filled circles correspond to different definitions of the interactions between words in sentences. The solid line \protect\cite{dm01f} shows the 
result of our calculations using the known parameters of the Word Web, namely 
the size $t \approx 470\, 000$ and the average number of connections, $\overline{k}(t) \approx 72$. The arrows indicate the theoretically obtained point of crossover, $k_{cross}$, between the regions with exponents $3/2$ and $3$, and the cutoff $k_{cut}$ of the power-law dependence due to finite-size effect. }
	 \label{f7}
\end{vchfigure}


In Ref. \cite{fs01}, the Word Web was constructed after processing $3/4$ million words of the British National Corpus. The British Corpus is a collection of text samples of both spoken and written modern British English. The resulting network contains about $470\,000$ vertices. The average degree is 
$\overline{k} \approx 72$. These are the only parameters of the network we know and can use in the model. 

Notice that the quality of the empirical data is \cite{fs01} is high: the range of degrees is five decades. The empirical degree distribution has two power-law regions with exponents $1.5$ and about $3$ (the latter value is less precise, 
 since statistics in this region is worse). The crossover point and the cutoff due to finite-size effect can be easily indicated (see Fig. \ref{f7}). 

We treat language as a growing network of interacting words. At its birth, a new word already interacts with several old ones. New interactions between old words emerge from time to time, and new edges emerge. All the time a word lives, it enters in new ``collaborations''. Therefore the number of connections grows more rapidly than the number of words: the growth of the Word Web is accelerated. 

How do words find their collaborators in language? Here we again use the idea of preferential linking \cite{ba99}, again the 
principle ``{\em popularity is attractive}'' works.  


\begin{vchfigure}[htb]
\includegraphics[scale=0.8]{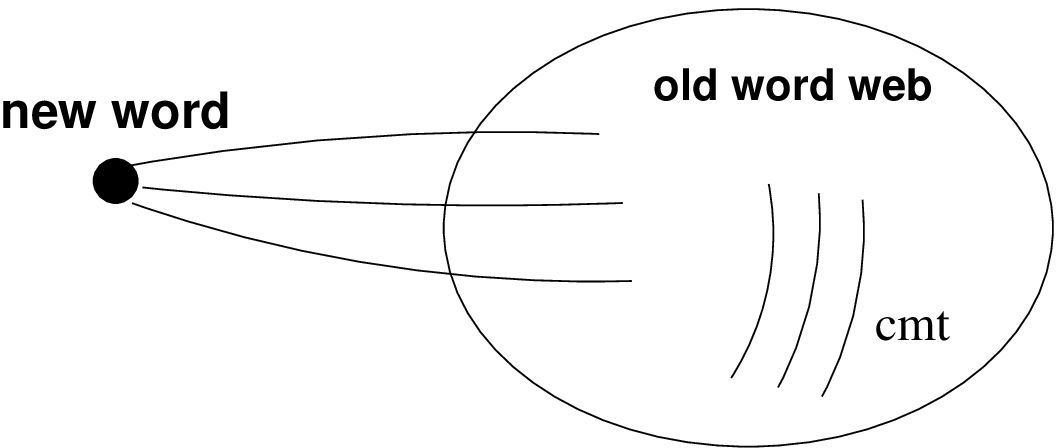}
	 \vchcaption{Scheme of the Word Web growth. At each time step a new word emerge, so that $t$ is the total number of words. It connects to $m \sim 1$ preferentially chosen old words. Simultaneously $cmt$ new edges emerge between pairs of preferentially chosen old words. 
We use the simplest rule of the preferential attachment when a node is chosen with the probability proportional to the number of its connections.}
	 \label{f8}
\end{vchfigure}


We use the following rules of the network growth (see Fig. \ref{f8}) \cite{dm01f}. 

(1) At each time step, a new vertex (word) is added to the network, and the total number of words is $t$. 

(2) At its birth, a new word connects to several old ons. Let, in average, this number be $m$, so that this number is not necessary integer. We use the simplest natural version of preferential linking: a new word become connected with some old one $i$ with the probability proportional to its degree $k_i$, like in the Barabasi-Albert model \cite{ba99}. 

(3) In addition, $cmt$ new edges emerge between old words, where $c$ is a constant coefficient that characterizes a particular network. 
If each vertex makes new connections with a constant rate, this linear dependence on time naturally arises. These new edges emerge between old words $i$ and $j$ with the probability proportional to the product of 
their degrees $k_ik_j$ \cite{dm00b}. 

These simple rules define the minimal model that can be solved exactly. Here we discuss only the results of the continuum approach. 
In this case, the approach gives an excellent description of the degree distribution and the proper values of exponents. 

In the model that we discuss, words are actually considered as collaborators in language. In our approach the essence of the evolution of language is the evolution of collaborations between words. Therefore the situation for the Word Web should be rather similar to that for networks of collaborations. The equivalent model was applied to scientific collaboration nets \cite{bjnr01a}, but the more complex nature of these networks makes the comparison impossible.     

As above, in the continuum approximation, we can write the equation for the average degree at time $t$ of the word that emerged at time $s$: 

\begin{equation}
\label{25}
\frac{\partial \overline{k}(s,t)}{\partial t} = 
(m + 2 cmt) \frac{\overline{k}(s,t)}{\int_0^t du\,\overline{k}(u,t)}
\, ,  
\end{equation} 
where the initial condition is $\overline{k}(0,0)=0$ 
and the boundary one is $\overline{k}(t,t)=m$.

One can see that the total degree of the network is 
$\int_0^t du\,\overline{k}(u,t) = 2mt + cmt^2$, so that its average degree 
at time $t$ is equal to $\overline{k}(t) = 2m + cmt$. 

The solution of Eq. (\ref{25}) is of a singular form 

\begin{equation}
\label{26}
\overline{k}(s,t) = m \left( \frac{cmt}{cms} \right)^{1/2} 
\left( \frac{2m+cmt}{2m+cms} \right)^{3/2}
\, .  
\end{equation}
The form of this equation indicates the presence of two distinct regimes in this problem. Using Eqs. (\ref{18}) and (\ref{26}) readily yields the non-stationary degree distribution  

\begin{equation}
\label{27}
P(k,t) = \frac{1}{ct} \frac{cs(2+cs)}{1+cs} \frac{1}{k} 
\, ,  
\end{equation}
where $s=s(k,t)$ is the solution of Eq. (\ref{26}). 
Notice that, formally speaking, the number $m$ is absent in Eq. (\ref{27}). 
This is the consequence of our definition of the coefficient 
$cm$ (see above).  

From Eqs. (\ref{26}) and (\ref{27}), one sees that the non-stationary degree distribution has two regions with different behaviors separated by the crossover point 

\begin{equation}
\label{28}
k_{cross} \approx m \sqrt{ct}(2+ct)^{3/2}  
\, .  
\end{equation}
The crossover moves in the direction of large degrees as the network grows. Below this point, the degree distribution is stationary, 

\begin{equation}
\label{29}
P(k) \cong \frac{\sqrt{m}}{2} k^{-3/2} 
\, .  
\end{equation} 

Above the crossover point, we obtain the behavior 

\begin{equation}
\label{30}
P(k,t) \cong \frac{(2m + cmt)^3}{4} k^{-3}
\, .  
\end{equation}
so that the degree distribution is non-stationary in this region. Thus, we have obtain two distinct values for the degree distribution exponent, namely, $3/2$ and $3$.    

The model that we consider has two limiting cases. When $c=0$, it turns to be the Barab\'asi-Albert model, where $\gamma=3$. When $m$ is small but $cm$ is large, we come to the network from Sec. \ref{ss-less} which has $\gamma=3/2$ and a stationary degree distribution. Thus these two values of $\gamma$ are not surprising. The important point is that the crossover is observable even though $cmt \gg m$. 

The degree distribution has one more important point, the cutoff produced by finite-size effect. We estimate its position from the condition 
$t \int_{k_{cut}}^\infty dk P(k,t) \sim 1$ (see Secs. \ref{s-degree} and \ref{s-general}). This yields  

\begin{equation}
\label{31}
k_{cut} \sim \sqrt{\frac{t}{8}} (2m+cmt)^{3/2}
\, .  
\end{equation} 

Using Eqs. (\ref{28}) and (\ref{30}) one can estimate the number of words above the crossover: 

\begin{equation}
\label{32}
N_{c} \approx t \int_{k_{cross}}^\infty dk P(k,t) \sim \frac{m}{8c}
\, .  
\end{equation} 

We know only two parameters of the Word Web that was constructed in Ref. \cite{fs01}, namely $t=0.470 \times 10^6$ and $\overline{k}(t) = 72 = 2m + cmt \approx cmt$. 
About $m$ we know only that it is of the order of $1$. 
From the above relations, one sees that the dependence on $m$ is actually weak and is not noticeable in log-log-scale plots.  
In fact, $m$ is inessential parameter of the model.  
Hence we can set its value to $1$. 

In Fig. \ref{f7}, we plot the degree distribution of the model (the solid line). To obtain the theoretical curve, we used Eqs. (\ref{26}) and (\ref{27}) with $m=1$ and $c \approx \overline{k}(t)/t$. A rather inessential deviations from the continuum approximation are accounted for in the small-degree region ($k \sim 10$). One sees that the agreement with the empirical data \cite{fs01} is fairly good. Note that we do not used any fitting. However, for a better comparison, in Fig. \ref{f7}, the theoretical curve is displaced upwards. Actually, this is not a fitting, since we have to exclude two empirical points with the smallest degrees. These points are dependent on the method of the construction of the Word Web, on specific grammar, so that any comparison in this region is meaningless in principle. 

From Eqs. (\ref{28}) and (\ref{31}), we find the characteristic values for the crossover and cutoff, $k_{cross} \approx 5.1 \times 10^3$, that is, 
$\log_{10}k_{cross} \approx 3.7$, and $\log_{10}k_{cut} \approx 5.2$. 
From Fig. \ref{f7} we see that these values  coincide with the experimental ones. 
We should emphasize that the extent of agreement is truly surprising. The minimal model does not account for numerous, at first sight, important factors, e.g., the death of words, the variations of words during the evolution of language, 
etc. 

The agreement is convincing since it is approached over the whole range of values of $k$, 
that is, 
over five decades. 
In fact, the Word Web turns out to be very convenient in this respect since the total number of edges in it is extremely high, about $3.4\times10^7$ edges, and the value of the cutoff degree  
is large. 

Note that few words are in the region above the crossover point $k_{cross} \approx 5.1\times 10^3$. 
These words have a different structure of connections than words from the rest part of language.   
With the growth of language, $k_{cross}$ increases rapidly but, as it follows from Eq. (\ref{32}), the total number $N_c$ of words of degree greater than $k_{cross}$ does not change. It is a constant of the order of 
$m^2/(8cm) \sim 1/(8c) \approx t/(8\overline{k}) \sim 10^3$, that is of the order of the size of 
a small set of words forming the 
kernel lexicon of the British English which was estimated as $5,000$ words \cite{fs00} and is the most important core part of language. Therefore, our concept suggests that the number of words in this part of language does not depend essentially of the size of language.  
Formally speaking, 
the size of this core determined by the value of the average rate $c$ with which words find new partners in language. 

If our simple theory of the evolution of language is reasonable then the sizes of the cores of primitive languages are close to those for modern ``developed'' languages.


\section{Wealth distribution in evolving societies}\label{s-wealth}

Ideas from network science can be applied to various problems. Here we show how the idea of the nonlinear growth works in econophysics. 

One of the basic problems of econophysics is wealth distribution. Usually, wealth distribution is treated by using so called stochastic multiplicative models. The standard description of these stochastic multiplicative processes is provided by the generalized Lotka-Volterra equation \cite{sl96,sc97}. The preferential linking mechanism, that is, the general ``{\em popularity is attractive}'' principle, provides the stochastic multiplicative dynamics of networks \cite{dm01e}. 
Therefore, results that were obtained for networks may be easily interpreted in terms of wealth distribution. 

Let us discuss briefly wealth distribution in stable (stagnating), developing, and degrading (dying) societies. 
For simplicity, in our very schematic consideration, we do not account for mortality, redistribution and loss of money, inflation, and many other important factors. Let there be one birth per time step. Therefore, there are $t$ members of the society at time $t$. Thus we consider growing (non-equilibrium) societies. 

In stable societies, wealth per member (the average capital, the average amount of money) does not change with time, and the input flow of capital is constant. 
In developing societies, the average wealth and the input flow of capital grow with time. In degrading societies, these quantities decrease.  

One introduces the distribution function of wealth, $P(k,t)$. 
If this distribution is a power law, $P(k) \sim k^{-\gamma}$, and $\gamma<2$, the society is {\em ``unfair''}: few persons keep a finite fraction of the total wealth. 
If $\gamma>2$, the society is {\em ``fair''}. The wealth condensation transition \cite{bm00} occurs when $P(k)$ passes over the $k^{-2}$ dependence. 
When $P(k)$ decreases more rapidly than a power law, e.g., the function is exponential, the society is {\em ``superfair''}.  

To study wealth distribution in various societies, we consider the simplest demonstrating case of a power-law input flow of capital $t^\alpha$. Growth exponent $\alpha$ indicates the type of society. $\alpha=0$ corresponds to stable societies. Positive and negative $\alpha$ exponents provide developing and degrading societies, respectively.  

Let us discuss the simplest situation. 
We assume that money attract money. 
While trying to diminish inequality, society permanently distribute some fraction of wealth ``fairly'' (equally) among its members. 
Another way to make life better for all is to provide everybody 
with a starting capital. 
Society also provides its members by the educational etc. ``capital'' which can also attract money. Such a factor, additional attractiveness, $A$, also proportional to the average wealth.  
It may be provided only once, at the birth, $A(s)$ ($s<t$ is the birth time of an individum), it may increase equally for all persons, $A(t)$ ($t$ is the age of a society), but in both cases the effect is qualitatively similar to starting capital. We consider the first possibility, that is, the providing with some starting capital at the birth as the simplest. 
 
We again apply the continuum approach. Then $\overline{k}(s,t)$ is the average wealth of the person that was born at time $s<t$, $t$ is the present time.


\subsection{Stable (stagnating) societies}\label{s-stable}

Let $m_s$ be starting capital and $m$ extra wealth be distributed at each time step. $A$ is a constant additional attractiveness. The total input flow of wealth is equal to $m+m_s$. A fraction $p$ of the flow $m$ is distributed among members of the society randomly, that is, ``fairly'', the flow $(1-p)m$ is distributed preferentially with probability proportional to your wealth. The continuum approach equation for the average individual wealth $\overline{k}(s,t)$ is of the form 

\begin{equation}
\label{33}
\frac{\partial \overline{k}(s,t)}{\partial t} = 
\frac{p m}{t} + (1-p)m 
\frac{\overline{k}(s,t)+A}{\int_0^t du\,[\overline{k}(u,t)+A]}  
\end{equation}
with the initial condition $\overline{k}(0,0)=0$ and the boundary one $\overline{k}(t,t)=m_s$. Integrating (by parts) both the sides of Eq. (\ref{33}) over $s$ yields naturally $\int_0^t ds\, \overline{k}(s,t) = (m+m_s)t$. Similarly to the calculations of Sec. \ref{ss-less}, we obtain the power-law wealth distribution with $\gamma$ exponent  

\begin{equation}
\label{34}
\gamma = 2 + \frac{pm+m_s+A}{(1-p)m} > 2
\, .
\end{equation} 
Thus, in stable societies, $\gamma>2$, so that a stagnating society is fair.


\subsection{Developing and degrading societies}\label{s-developing}

Here we discuss a natural case: let your starting capital be proportional to the average wealth in the society at your birth, $m_s(t) = dmt^\alpha$, where $d$ is a positive constant. In addition, the wealth $mt^\alpha$ is distributed among members of the society at each increment of time. The wealth $pmt^\alpha$ is distributed equally. The wealth $(1-p)mt^\alpha$ is distributed 
 preferentially (money come to money).  
For brevity, we set $A(s,t) = 0$. 
Then we have 

\begin{equation}
\label{35}
\frac{\partial \overline{k}(s,t)}{\partial t} = 
m t^\alpha \frac{p}{t} +
(1-p)m t^\alpha
\frac{\overline{k}(s,t)}{\int_0^t du\,(u,t)} 
\, .
\end{equation} 
The initial and boundary conditions are $\overline{k}(0,0)=0$ and $\overline{k}(t,t)=dmt^\alpha$, respectively. From Eq. (\ref{35}), one sees that 
$\int_0^t ds\, \overline{k}(s,t) = m(1+d)t^{\alpha+1}/(\alpha+1)$. 

From Eq. (\ref{35}) we obtain the wealth distribution for various values of the parameters of the problem, $p$, $d$, and $\alpha$. 

(i) When 
$\alpha > (1-p)/(p+d)$, or, in other words, $p>(1-\alpha d)/(1+\alpha)$, 
the wealth distribution is exponential (the ``superfair society''). 

(ii) For $\alpha < (1-p)/(p+d)$, we obtain the power-law wealth distribution with exponent   

\begin{equation}
\label{36}
\gamma = 2 + \frac{(1+\alpha)(p+d)}{1 - p - \alpha(p+d)}  
\, .
\end{equation} 
One sees that $\gamma=2$ at $\alpha=-1$. This corresponds to the ``wealth condensation transition'' from the ``fair'' society ($\gamma>2$ for $\alpha>-1$) to the ``unfair'' one ($\gamma<2$ for $\alpha<-1$).  
The resulting phase diagram is shown in Fig. \ref{f9}. 
Note that the position of the wealth condensation transition does not depend on  particular values of $p$ and $d$. 
Therefore, even if a significant part of new wealth  
is distributed equally, rapidly degrading societies are necessarily unfair! 


\begin{vchfigure}[htb]
\includegraphics[scale=0.8]{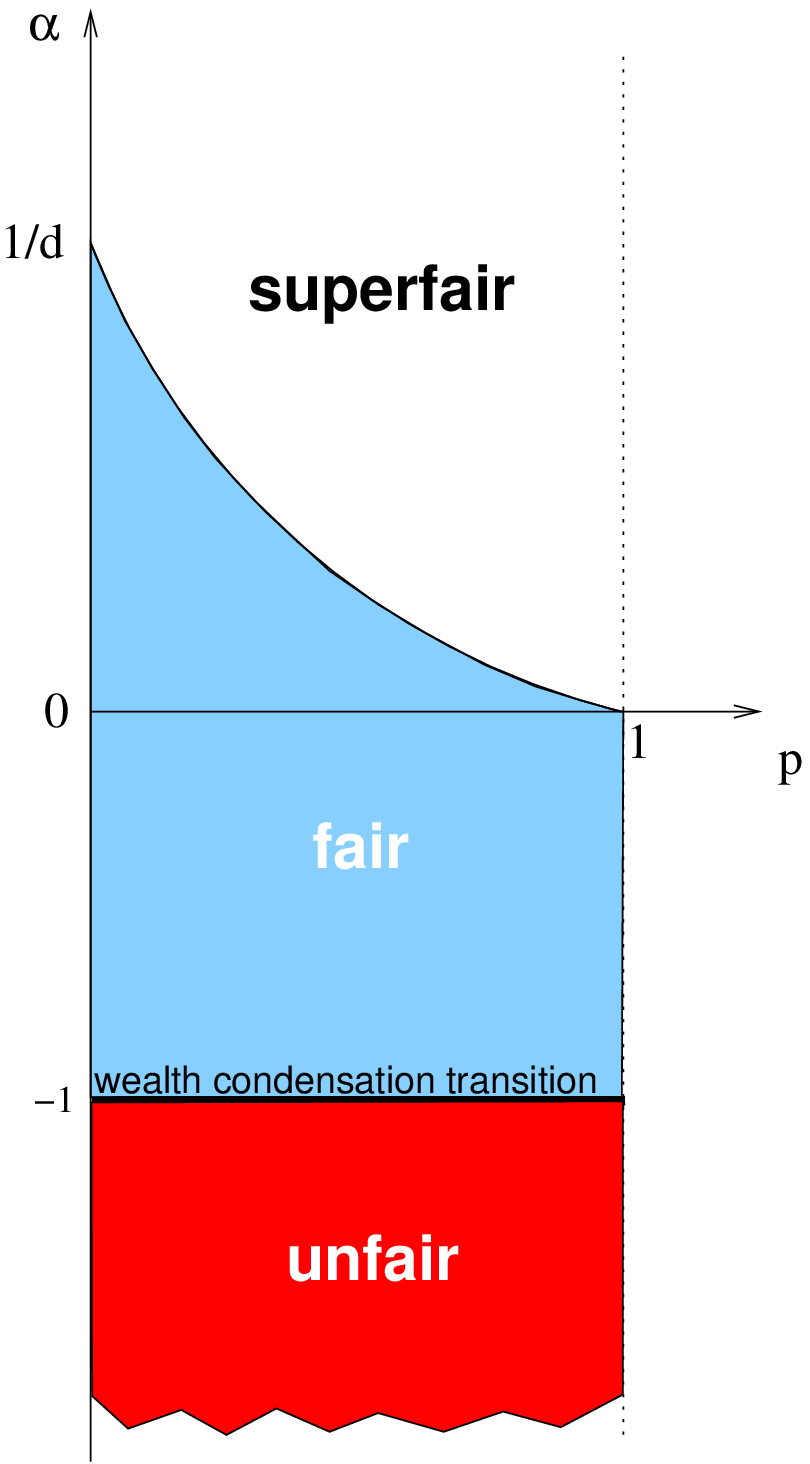}
	 \vchcaption{Phase diagram of evolving societies. 
At each time step, the wealth $pmt^\alpha$ is distributed equally between all members of a society, 
the wealth $(1-p)mt^\alpha$ is distributed preferentially (money come to money). The individual 
starting capital is $dmt^\alpha$. 
In our schematic model, the wealth distribution of the ``superfair'' society is exponential. 
The wealth distribution of the fair society is a power-law with exponent $\gamma>2$. For the 
unfair society $\gamma<2$.}
	 \label{f9}
\end{vchfigure}

 
The general picture of wealth distribution in our minimal approach is quite  
natural. Extremely degrading societies are unfair. It is impossible to approach any ``fairness'' by the ``fair'' distribution of any part of new wealth in such a situation. ``Fair'' societies are possible only if there is some progress or the degradation is rather modest. Only in fair societies ``fair'' distribution of new wealth produces visible results.


\section*{CONCLUSIONS}

The nonlinear growth of networks is more general situation than the linear growth. In real evolving networks, the nonlinear, in particular, accelerated growth is widespread and is the rule and not the exception. In many cases, it is impossible to understand the nature of an evolving network without accounting for this acceleration. 

The complicating circumstance is that existing empirical data clearly indicate the presence of the acceleration but usually fail to yield its quantitative description. Theoreticians may easily choose any functional form for the non-linear growth, but do these beautiful dependences have any relation to reality?


\section*{ACKNOWLEDGMENTS}

S.N.D. thanks PRAXIS XXI (Portugal) for a research grant PRAXIS XXI/BCC/16418/98. S.N.D. and J.F.F.M. were partially supported by the project POCTI/99/FIS/33141. We also thank A.V.~Goltsev and A.N.~Samukhin for many useful 
discussions\vspace{10pt}. 
\\ 
$^{\ast}$      Electronic address: sdorogov@fc.up.pt\\
$^{\dagger}$   Electronic address: jfmendes@fc.up.pt

\renewcommand\bibname{Bibliography}

\end{document}